\documentclass[twocolumn,showpacs,showkeys,aps,amsfonts,amssymb ]{revtex4-1}

\usepackage{graphicx}  
\usepackage{amsmath, amssymb, bm,bbm}   
\usepackage[dvips]{color}
\usepackage{natbib}
\usepackage{hyperref}
\usepackage{slashed}

\usepackage{longtable}

\usepackage{mathtools}

\def\){\right)} 
\def\({\left(} 
\def\]{\right]} 
\def\[{\left[}

\def\intA{$1\gtrsim g_t\gtrsim 0.55$}
\def\intB{$0.55\gtrsim g_t\gtrsim 0.35$}
\def\intC{$0.35\gtrsim g_t\gtrsim 0.1$}

\def\nopi{$\slashed{\pi}$EFT}

\newcommand{\simge}{\hspace*{0.2em}\raisebox{0.5ex}{$>$}
     \hspace{-0.8em}\raisebox{-0.3em}{$\sim$}\hspace*{0.2em}}

\begin{document}

\title{Fate of the neutron-deuteron virtual state as an  Efimov level}

\author{%
Gautam Rupak}
\email{grupak@ccs.msstate.edu}

\author{%
Akshay Vaghani}
\email{av298@msstate.edu}

\address{Department of Physics \& Astronomy and HPC$^2$ Center for Computational Sciences, 
Mississippi State
University, Mississippi State, MS 39762, USA}

\author{%
Renato Higa}
\email{higa@if.usp.br}
\address{
 Instituto de F\'\i sica, Universidade de S\~ao Paulo, 
 R. do Mat\~{a}o 1371, 05508-090, S\~ao Paulo, SP, Brazil
}

\author{%
Ubirajara van Kolck}
\email{vankolck@ipno.in2p3.fr}
\address{
Institute de Physique Nucl\'eaire, CNRS/IN2P3,
Universit\'e Paris-Sud, Universit\'e Paris-Saclay, 91406 Orsay, France}
\address{
Department of Physics, University of Arizona, 
Tucson, AZ 85721, USA}

\begin{abstract}
The emergence of Efimov levels in a three-body system is investigated near the unitarity limit characterized by resonating two-body interaction. No direct evidence of Efimov levels is seen in the three-nucleon system since the triton is the only physical bound state. We provide a model-independent analysis of nucleon-deuteron scattering at low energy by formulating a consistent effective field theory. We show that virtual states evolve into shallow bound states, which emerge as excited triton levels as we drive the system towards unitarity. Even though we consider this specific system, our results for the emergence of the Efimov levels are universal.
 \end{abstract}

\maketitle

\section{Introduction}
\label{sec:intro}

A distinctive feature of three-body physics is the Efimov 
effect~\cite{Efimov:1970,Efimov:1971,Amado:1971,Amado:1972}. 
In the presence of resonant 
two-body scattering, the three-body system supports
a tower of bound states whose binding energies display 
geometric scaling. 
The number of three-body bound states 
was predicted to scale 
as $\ln(a_0/r_0)$, where $a_0$ is the resonating two-body $s$-wave scattering 
length, and $r_0\ll a_0$ is the range of the interaction. 
Experimental verification through the detection 
of an excited state was provided three decades later in cold-atom 
experiments~\cite{Kraemer:2006, Knoop:2009,Zaccanti:2009,Gross:2009},
where the ratio $a_0/r_0$ could be increased by varying magnetic fields
near a Feshbach resonance.

Recently, the first excited Efimov state was identified \cite{Kunitski:2015qth} for atomic $^4$He, where $a_0/r_0$ is sufficiently large even in the absence of external magnetic fields. In complex nuclei the ratio $a_0/r_0$ is often unknown, but in some cases it might be large enough to accommodate an excited state: halo nuclei such as $^{11}$Li, $^{22}$C, or even as heavy as $^{62}$Ca, where two neutrons are weakly bound to a tight nuclear core, provide opportunity to observe the Efimov effect~\cite{Jensen:2004zz,Frederico:2012xh,Hammer:2017tjm}.

Neutron-deuteron ($nd$) at low energy is another system that 
could provide evidence of Efimov physics.   
Low-energy scattering is dominated by $s$ waves 
with spins $S=3/2$ (quartet) and  
$S=1/2$ (doublet). 
In the quartet channel,  
where spins are aligned, the Pauli exclusion principle  
prevents a 
three-nucleon (3N) bound state. 
In the doublet channel, the strong nuclear force leads to an attractive 
interaction that supports a  
3N bound state, the triton ($^3$H) with a binding energy $B_3\simeq 8.48$ MeV.
The next deeper Efimov state would appear only at $\sim 4$ GeV,
beyond the regime where a description in terms of nucleons
makes sense.
Here we search for a remnant of the next {\it shallower} Efimov state.

The low-energy phase shift in the doublet channel was analyzed by van Oers 
and Seagrave~\cite{vanOers:1967}, who suggested a modified effective range 
expansion (ERE) to describe the data below the deuteron breakup momentum.  
The presence of a virtual state with binding energy $\simeq 0.515$ MeV
was inferred~\cite{GirardFuda:1979}.
Adhikari {\it et al.}~\cite{Adhikari:1982} 
showed in a separable potential model that this virtual state is related to the 
Efimov spectrum.  
A similar connection for $^{20}$C \cite{Yamashita:2007ej,Yamashita:2008sg}
and for atomic systems \cite{Shalchi:2017gjl}
was investigated within a
zero-range model.

In contrast, we use the effective field theory (EFT) formalism to provide a 
model-independent analysis of the $nd$ system at low energies.  
All interactions allowed by  
symmetries are constructed with the relevant low-energy degrees of freedom, 
without modeling the high-energy physics. 
A  systematic scheme for calculations is formulated by expressing 
observables as an expansion in the small ratio 
$p/\Lambda_b$, where
$p$ is a typical low momentum scale associated with the processes
of interest
and $\Lambda_b$ is a high momentum scale that marks the breakdown
of the EFT.

In the so-called pionless EFT (\nopi) the relevant degrees of freedom are 
nonrelativistic nucleons 
(and other light particles such as photons, electrons and neutrinos)
with $\Lambda_b\sim m_\pi$, the pion mass, 
associated with the pion physics that is not included explicitly.
\nopi~has been very successful in describing 
two- and three-nucleon systems --- see for example 
Refs.~\cite{Chen:1999tn,Kong:1999sf,Bedaque:1999ve,Rupak:1999rk,Hammer:2001gh,Bedaque:2002yg,
Vanasse:2013sda,Konig:2015aka} --- when the typical momentum is
taken to be $p\sim\gamma\sim a_t^{-1}$, 
with $\gamma\simeq 45.7$ MeV the deuteron binding momentum
and $a_t\simeq 5.4$ fm the scattering length in the two-nucleon 
(2N) $^3S_1$ channel.

EFT enables us to study $nd$ scattering  
in the limit $a_t\to \infty$,
where the 2N scattering amplitude is only bounded by unitarity. 
We first fit the parameters of \nopi ~to reproduce the relevant 
2N and 3N experimental data. 
\nopi ~is then treated as the 
underlying theory that is used to generate 
artificial ``data'' at increasingly large values of $a_t$.
We develop a new low-energy EFT, which we refer to as halo EFT, 
to provide a model-independent analysis of the 
3N phase shift at low momentum generated from \nopi. 
This halo EFT, which treats the deuteron as an
``elementary'' particle and is thus applicable only 
below the deuteron breakup, is modeled after other halo EFTs 
where different clusters of nucleons are treated as 
relevant degrees of freedom \cite{Bertulani:2002sz,Bedaque:2003wa}.
Here, the deuteron is the core and the neutron forms the ``halo'' around it, 
consisting of shallow virtual and bound states. 
The halo EFT provides a theoretical basis for the modified ERE 
obtained empirically by van Oers and Seagrave,
and allows us to track the virtual state at unphysical scattering lengths.
As we drive \nopi~towards the unitarity limit, 
the binding energy of the virtual state 
decreases till it becomes the first excited bound state of the triton,
thus demonstrating its Efimov character.
Higher Efimov states appear in the same way if the scattering lengths are
increased further. 
Although we focus on the 3N system, 
our framework could be applied to study the emergence of Efimov levels
in other systems as well.

\section{Pionless EFT}
\label{sec:nopi}

The  doublet $nd$ scattering amplitude  
was first calculated at leading order (LO) in \nopi ~in 
Ref.~\cite{Bedaque:1999ve}. 
It receives contributions from 
the LO
2N interactions, which consist of a single non-derivative contact
operator in each 2N $s$ wave
($^3S_1$  and $^1S_0$) 
with strength determined in terms of the respective 
scattering length.
In addition, there is a contribution from a 
3N non-derivative contact interaction, which is needed to 
render the amplitude well-defined.
Next-to-leading-order (NLO) corrections introduce a two-derivative 
interaction in each 2N $s$ wave, which can be constrained by the 
corresponding effective range. 
No new 
3N
interaction contributes at this order~\cite{Bedaque:1999ve,Hammer:2001gh}. 
A  
momentum-dependent 
3N interaction enters at NNLO~\cite{Bedaque:2002yg}. 
To this order,
the EFT expansion has been shown to be convergent 
\cite{Vanasse:2013sda}, and to reproduce both
experimental data (when available) and results
from sophisticated phenomenological potentials.  

The unitarity limit corresponds 
not only to arbitrarily large 2N scattering lengths  
but also to vanishing 2N effective ranges and higher ERE parameters.
This removes 
higher-order corrections in the 
2N interaction, so that a LO calculation is sufficient to explore
the connection to the Efimov spectrum. 
The LO $nd$ $T$-matrix 
$T_t(p)$ is obtained from two coupled integral equations~\cite{Bedaque:1999ve}:
\begin{align}\label{eq:int_equation_triton}
\begin{multlined}A_t(k,p) =  \frac{1}{4}\Big[\mathcal{K}(k,p) +h_0(\lambda)\Big]\nonumber\\
+\int_{0}^{\lambda} \frac{dq q^2}{2\pi} \Big\{
\Big[\mathcal{K}(k,q)+h_0(\lambda)\Big]\mathcal{D}_t(E;q)A_t(q,p) \nonumber
\\
+\Big[3\mathcal{K}(k,q)+h_0(\lambda)\Big]\mathcal{D}_s(E;q)A_s(q,p)
\Big\}
\, , \nonumber
\end{multlined}\\
\begin{multlined}
A_s(k,p) =  \frac{1}{4}\Big[3\mathcal{K}(k,p) +h_0(\lambda)\Big]\nonumber\\
+\int_{0}^{\lambda} \frac{dq q^2}{2\pi} \Big\{
\Big[3\mathcal{K}(k,q)+h_0(\lambda)\Big]\mathcal{D}_t(E;q)A_t(q,p)
\nonumber \\
+\Big[\mathcal{K}(k,q)+h_0(\lambda)\Big]\mathcal{D}_s(E;q)A_s(q,p)
\Big\}\nonumber\, , 
\end{multlined}\\
\mathcal K(k,q)=\frac{1}{kq}\ln\left(\frac{k^2+k q +q^2-m_N E- i0^+}
{k^2-k q+q^2-m_N E-i0^+}\right)\, ,
\end{align}
where $E=3p^2/(4 m_N)-\gamma^2/m_N$ is the total center-of-mass energy
and $m_N\simeq 938.9$ MeV is the isospin-averaged nucleon mass.  
The intermediate 2N 
contribution is contained in the two-point Green's function 
$\mathcal D_\eta(E;p)=(-\gamma_\eta+\sqrt{-m_N E +p^2/{4}-i 0^+})^{-1}$, 
where $\eta=t$ ($s$) for the $^3S_1$ ($^1S_0$) channel.
In the $^3S_1$ channel we use the deuteron binding momentum to set 
$\gamma_t=g_t\gamma$, 
and in the $^1S_0$ channel, where no bound state exists, 
we use the scattering length $a_s\simeq -23.714$ fm to fix 
$\gamma_s=g_s/a_s$, which
is consistent with the LO power counting~\cite{vanKolck:1998bw,Chen:1999tn}. 
The physical point corresponds to $g_t=g_s=1$.
The $3N$ amplitudes $A_{t,s}=m_NT_{t,s}/(3\pi)$ do not converge when  
the regulator is removed, $\lambda\rightarrow\infty$,
unless we fix the 3N parameter $h_0(\lambda)$ to guarantee 
that one 3N observable is made regulator independent. 
For any given cutoff $\lambda$, we tune $h_0$ such that we reproduce the 
doublet scattering length $a_3=0.65$ fm~\cite{Dilg:1971}
for $g_t=g_s=1$. This determines the independent
parameter $\Lambda_\star$ 
appearing in the log-periodic 3N force \cite{Bedaque:1999ve}.

\begin{figure}[tb]
\begin{center}
\includegraphics[width=0.49\textwidth,clip=true]{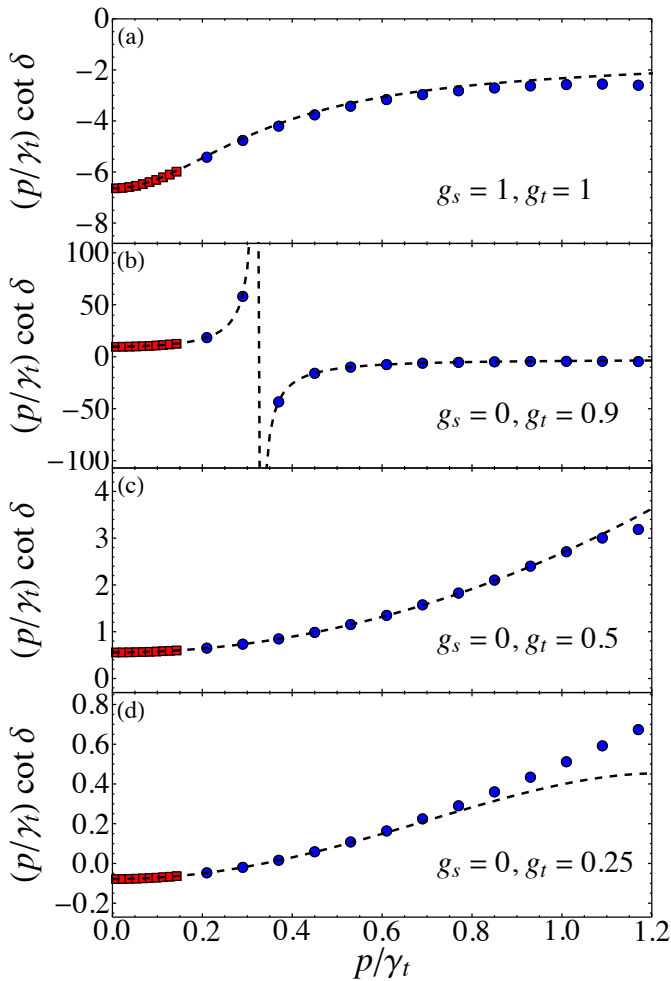} 
\end{center}
\caption{Phase shift: $p\cot\delta$ 
as function of the momentum $p$,
both in units of the deuteron binding momentum $\gamma_t$.
The (blue) dots and (red) squares are numerical results from solving 
Eq.~(\ref{eq:int_equation_triton}) 
for several values of $g_t$, $g_s$ as indicated,
once the 3N parameter $\Lambda_\star$ is determined by
imposing $a_3=0.65$ fm for $g_t=g_s=1$. 
The (black) dashed curves are modified ERE fits to the data over a small momentum 
range indicated as (red) squares. 
We expect the modified ERE fits to describe the numerical results 
for $p/\gamma_t \lesssim 2/\sqrt{3}$. }
\label{fig:pcotdelta}
\end{figure}
The doublet phase shift is obtained from
$p\cot\delta=ip+1/A_t(p,p)$.
In Fig.~\ref{fig:pcotdelta}, we show the phase shift calculated from 
the three-body 
integral equations in Eq.~(\ref{eq:int_equation_triton}) at the physical
point and 
a few results as we approach the unitarity limit.  
The only physical parameters that enter the microscopic calculation using 
\nopi~are $g_t$, $g_s$, and $\Lambda_\star$. 
We approach unitarity taking $g_s=0$ and 
making the deuteron arbitrarily shallow,
$g_t\rightarrow 0$, while keeping $\Lambda_\star$ fixed.

For short-ranged interactions, $p\cot\delta$ is an analytic function of $p^2$. 
However, $p\cot\delta$ rises rapidly at low momenta --- see panel (a)  
of Fig.~\ref{fig:pcotdelta} --- and for $g_t=g_s=1$ a simple Taylor series 
expansion around $p=0$ gives a poor description even at relatively small 
momenta $p\sim 10$ MeV. Instead, the modified ERE~\cite{vanOers:1967}
\begin{align}\label{eq:modifiedERE}
p\cot\delta &= \frac{-1/a+r p^2/2+ sp^4/4+\cdots}{1+ p^2/p_0^2} \nonumber\\
&=-\frac{R}{1+p^2/p_0^2}-A + B p^2 +\cdots\, ,
\end{align}
works remarkably well up to about the deuteron breakup momentum 
$2\gamma/\sqrt{3} \simeq 53$ MeV. 
While bound and virtual states correspond to poles of the $T$-matrix 
$T_t(p)$ at imaginary momenta $p=i\kappa_j$, 
there is also a pole in $p\cot\delta$ at $p^2=-p_0^2$ 
that corresponds to a zero of $T_t(p)$.
When we fit the modified ERE to the LO phase shift in 
panel (a) of Fig.~\ref{fig:pcotdelta},
we get the fit 
parameters $a\approx0.65$ fm $=a_3$, $r\approx-141$ fm, $s\approx 62$ fm$^3$, 
and $p_0\approx 16.1$ MeV,
which translate into a shallow virtual state 
at $\kappa_1\approx -26.8$ MeV 
with a binding energy $\approx$ 0.574 MeV. 
Naively one might expect all the scattering parameters to scale 
with some power of the range of nuclear interaction 
$\sim m_\pi^{-1}\simeq 1.4$ fm. 
The fitted parameters $p_0^{-1}$, $a^{-1}$ and $r$ are 
unusually large 
compared to the naive expectation. 
The standard ERE holds only 
for $p\ll p_0$, with large inverse scattering length $a^{-1}$,
effective range $r+2(ap_0^{2})^{-1}$, {\it etc.}

\section{Halo EFT and modified ERE}
\label{sec:haloEFT}

We now formulate the halo EFT that describes  
$nd$ scattering below the deuteron breakup momentum $2\gamma_t/\sqrt{3}$,  
and provides a justification for Eq. \eqref{eq:modifiedERE}.
In this theory the 
deuteron is treated as a fundamental particle, and so the breakup momentum 
sets the breakdown scale $\Lambda_b\sim \gamma_t$. 
A shallow $s$-wave pole by itself can be accounted for in the 
standard ERE by a large 
scattering length such as in the 
2N $^1S_0$ and $^3S_1$ channels,
requiring a fine-tuned interaction to be treated nonpertubatively at 
LO~\cite{vanKolck:1998bw,Chen:1999tn}.
A shallow amplitude zero by itself requires another fine-tuning 
that leads to a perturbative amplitude
with a large effective range and a small scattering 
length~\cite{vanKolck:1998bw}. 
Here we identify 
two momentum scales associated with the presence
of both zero and 
virtual pole, 
respectively $|p_0|\sim Q$ and $|\kappa_1|\sim \aleph$.
The parameters $a$, $r$ and $p_0$ are fine-tuned, and we require three 
fine-tuned couplings to reproduce the desired modified ERE.
The halo EFT is conveniently written using two auxiliary 
fields as
\begin{align}
\mathcal L =& n^\dagger\left(i\partial_0+\frac{\bm{\nabla}^2}{2m_N}\right) n
+
\vec{d}{\;}^\dagger\cdot \left(i\partial_0+\frac{\bm{\nabla}^2}{2m_d}\right) \vec{d}\nonumber\\
&+\sum_{j=1}^2{\psi^{(j)}}^\dagger\left[\Delta_j+c_j\left(i\partial_0
+\frac{\bm{\nabla}^2}{2 M}\right)\right] \psi^{(j)}\nonumber \\
&+\sqrt{\frac{2\pi}{3\mu}}
\left[\left({\psi^{(1)}}^\dagger+{\psi^{(2)}}^\dagger\right)
\vec\sigma\cdot n \vec{d}
+\operatorname{H.c.}\right] 
+\cdots\, .
\end{align}
Here, 
$n$ is the spin-doublet neutron field with mass $m_N$, 
$\vec{d}$ is the spin-triplet 
deuteron field with mass $m_d\approx 2 m_N$, and 
$\psi^{(j)}$ with $j=1,2$ are two auxiliary spin-1/2 fields 
with total mass $M=m_N+m_d\approx3 m_N$ and 
residual masses $\Delta_j$.
$\vec{\sigma}$ are Pauli matrices that act on the (suppressed) spinor indices 
of  $n$ and $\psi^{(j)}$. 
We chose to fix the couplings of both auxiliary fields to neutron and deuteron
in terms of the reduced mass $\mu=m_N m_d/M\approx 2 m_N/3$,
transferring their strength to parameters $c_j$~\cite{Griesshammer:2004pe}.
In the power counting discussed below,
$c_2\ll c_1$ and the corresponding interaction appears at subleading orders
together with the interactions lumped into the ``$\cdots$''.
Integrating out $\psi^{(2)}$ one recovers the Lagrangian 
from Ref. \cite{Kaplan:1996nv},
but the scaling of parameters is different here. 
Our approach inspired a reformulation of chiral EFT in
the 2N $^1S_0$ channel where not only the shallow virtual state
but also the amplitude zero is taken into 
account~\cite{SanchezSanchez:2017tws}.

A straightforward calculation of the $nd$ scattering amplitude $T_t(p)$ 
at LO in the halo EFT, where the contribution from the loops is resummed, gives
\begin{align}\label{eq:AmplitudeEFT}
iT_t(p)= \frac{2\pi i}{\mu}\Bigg[-\left(\frac{1}{\Delta_1\!+\!c_1 p^2/(2\mu)} 
+\frac{1}{\Delta_2} \right)^{-1}  
\!\!- L(p)\Bigg]^{-1}\!\! .\nonumber\\
\end{align}
The loop contribution,
\begin{align}
L(p)= 
4\pi\int \frac{d^3\bm{q}}{(2\pi)^3}\frac{1}{q^2-p^2-i 0^+}
=ip\, ,
\end{align}
was, for simplicity, evaluated in dimensional regularization using 
minimal subtraction. The final result is independent of the regularization 
method. 
We obtain the modified ERE (\ref{eq:modifiedERE}) 
from Eq.~(\ref{eq:AmplitudeEFT}) with
\begin{align}\label{eq:EFTcouplings}
p_0^2= &
\frac{2\mu}{c_1}\left(\Delta_1+\Delta_2\right)\, , &
\frac{1}{a}= &
A+R =\frac{\Delta_1 \Delta_2}{\Delta_1+\Delta_2}\, , \nonumber \\
-\frac{r}{2}= &
\frac{A}{p_0^2}=\frac{c_1}{2\mu}\,\frac{\Delta_2}{\Delta_1+\Delta_2}\, , & &
\end{align}
the shape-like parameter $s$ appearing at higher order.

\section{Analysis}
\label{sec:analysis}

To develop a consistent power counting for the halo EFT, 
we start with the analytic structure of the $T$-matrix 
$T_t(p)$. 
The $T$-matrix poles here are the roots of 
 \begin{align}\label{eq:rootrelation}
 p\cot\delta - ip =& \frac{-1/a+ rp^2/2}{1+ p^2/p_0^2}-ip \nonumber \\
=& -i\, \frac{(p-i\kappa_1)(p-i\kappa_2)(p-i\kappa_3)}{p^2+p_0^2}\, ,
 \end{align}
 with 
 \begin{align}\label{eq:rootsA}
                                           \kappa_1+\kappa_2+\kappa_3 = &-\frac{r}{2}\, p_0^2 = A\, ,  \nonumber\\
 \kappa_1\kappa_2+\kappa_2\kappa_3+\kappa_3\kappa_1=&-p_0^2\, , \quad  \kappa_1\kappa_2\kappa_3=-\frac{p_0^2}{a}\, .
 \end{align}
Using the parameters $a$, $r$, $p_0$ fitted earlier for $g_t=g_s=1$, 
the three roots $i\kappa_1\approx-27i$ MeV, $i\kappa_2\approx 35i$ MeV, 
and $i\kappa_3\approx 83i$ MeV are imaginary.
As we move towards unitarity ($g_s=0$, $g_t\rightarrow 0$),
we refit the \nopi~results with Eq. \eqref{eq:rootrelation},
as shown in Fig. \ref{fig:pcotdelta}. The roots remain imaginary.
The third root 
is always deeper than the 
breakdown scale $\Lambda_b\sim\gamma_t$ of the halo EFT, 
and is, therefore, not relevant.

\begin{figure}[tb]
\begin{center}
\includegraphics[width=0.49\textwidth,clip=true]{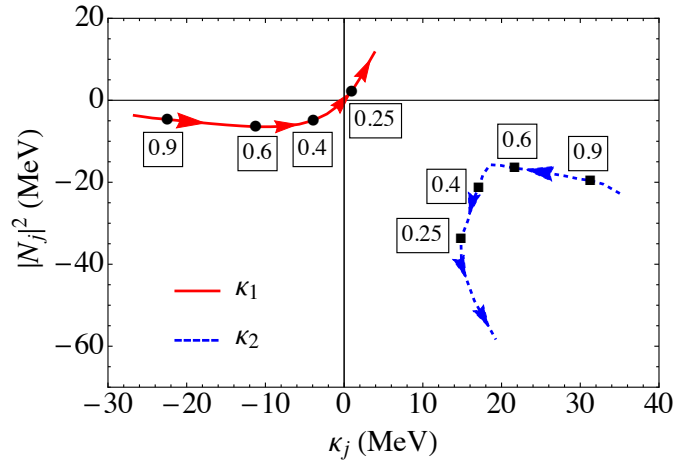} 
\end{center}
\caption{Normalizations $|N_j|^2$ for the imaginary roots $i\kappa_1$ 
(red solid curve) and $i\kappa_2$ (blue dashed curve). 
The arrows indicate the evolution of the states as we make $g_t$ smaller.
A few $g_t$ values are indicated as (black) circles on the $\kappa_1$ curve and 
as (black) squares on the $\kappa_2$ curve. 
}
\label{fig:residue}
\end{figure}

Near the poles we can write the $S$-matrix 
$S_t(p) =\exp[i2\delta(p)]\approx \sum_j R_j/(p-i\kappa_j)+f(p)$,
where $R_j$ are the residues at the poles, and $f(p)$ some finite piece. 
We provide an interpretation of the shallower roots based on the residues 
$R_{1,2}$.
The wavefunction normalizations for the two shallowest poles are
\begin{align}
|N_1|^2=iR_1=&\frac{2\kappa_1(\kappa_1^2-p_0^2)}{(\kappa_1-\kappa_2)(\kappa_1-\kappa_3)}\, , \nonumber\\
|N_2|^2=iR_2=&\frac{2\kappa_2(\kappa_2^2-p_0^2)}{(\kappa_2-\kappa_1)(\kappa_2-\kappa_3)}\, , 
\end{align}
and their evolution towards unitarity is shown in Fig. \ref{fig:residue}.
For a bound state, the normalization must be non-negative. 
At the physical point $g_t=g_s=1$, the shallowest root 
$i\kappa_1$ with $\kappa_1<0$ 
is a pole on the second 
Riemann energy sheet with a negative normalization $iR_1<0$, and thus 
describes the virtual state that has been identified in the 
past~\cite{GirardFuda:1979}. 
The second root $i\kappa_2$ is a pole on the first energy sheet since $\kappa_2>0$. 
However, its normalization is also negative, $iR_2<0$ --- it is 
a  ``redundant pole''~\cite{Ma:1947zz,EdenTaylor:1964}
and does not describe a physical state. 
As we take the limit $g_t\rightarrow0$ at $g_s=0$, 
the second root $i\kappa_2$ remains a redundant pole and the first 
root $i\kappa_1$ evolves from a virtual to a real bound state. 

In devising the power counting for the halo EFT, the three physical scales 
identified earlier, 
$|p_0|\sim Q$, $|\kappa_1|\sim\aleph$ and $\Lambda_b\sim \gamma_t$,
have to be taken into account.  The scale $Q$ starts as the smallest, gets 
smaller in size and then grows.  When $|p_0|$ grows beyond the 
breakdown scale, $|p_0|\simge \Lambda_b$, 
it stops being relevant in 
halo EFT. We find it convenient to separate the evolution into three 
intervals of $g_t$ to account for 
different relative sizes of $Q$. 
Moreover, as the modified ERE is most easily expressed in terms of the 
scattering parameters $a$, $r$, $p_0$ and $s$, we use these to write the 
power counting for the renormalized halo EFT couplings $c_1$, $c_2$, 
$\Delta_1$ and $\Delta_2$. 

Initially, for \intA, we have
$Q\ll\aleph\ll \Lambda_b$. 
The sizes of $|\kappa_1|$ and $|\kappa_2|$ are similar and we also identify 
$|\kappa_2|\sim\aleph$ to avoid introducing another scale. 
As we make $g_t$ smaller, we find that $\kappa_3$ gets deeper while
$\kappa_1$ gets shallower, so we take $\kappa_3\sim A\sim \Lambda_b^2/\aleph$.
These roots arise if the halo EFT couplings are large,
$\Delta_{1,2}\sim \Lambda_b^2/\aleph$
and $c_1/(2\mu)\sim\Lambda_b^2/\aleph^3$,
but with a cancelation in 
$\Delta_1+\Delta_2\sim \Lambda_b^2Q^2/\aleph^3\ll\Delta_{1,2}$.
From Eq. \eqref{eq:EFTcouplings} we see that $p_0$ comes out 
shallow as assumed, while $1/a\sim R \sim \aleph\Lambda_b^2/Q^2\gg A$
and $r\sim\Lambda_b^2/(\aleph Q^2)$ are large.
For $p\ll \aleph$, Eq.~(\ref{eq:rootrelation}) is dominated
by the $a$ term containing the amplitude zero,
and for $p\ll Q$ the effective range is $\sim 2(ap_0^2)^{-1}$. 
In contrast, for $p\simge \aleph$ 
the $r$ term and the unitarity term ($-ip$) become comparable
to the $1/a$ contribution and generate the $T$-matrix poles.
If we take $c_2/(2\mu)\sim Q^2/\Lambda_b^3\ll c_1/(2\mu)$,
then the shape-like parameter 
$s = -(c_1 c_2/\mu^2)/(\Delta_1 +\Delta_2)\sim 1/\Lambda_b^3$, which is
consistent with its fit value at $g_t=1$.
Its contribution 
for $p\sim\aleph$ is suppressed by a factor of 
$sp^2/r\sim Q^2\aleph^3/\Lambda_b^5\ll 1$ compared to 
the $r$ and $a$ contributions.
Other modified ERE parameters appear at even higher orders.
The halo EFT with the power counting we propose leads to a model-independent 
derivation of the modified ERE, and describes the data accurately.

As we make $g_t$ smaller, $p_0^2$ gets smaller and changes sign 
such that $p\cot\delta$ develops a pole at real momentum around $g_t=0.9$,
analogous to the Ramsauer-Townsend effect~\cite{ramtown1,ramtown2}.
This is shown in panel (b) of Fig.~\ref{fig:pcotdelta}. 
The halo EFT (and the modified ERE) still gives a good description of the 
phase shift through the pole in $p\cot\delta$ even though the EFT couplings 
are fitted at momentum below the pole, as indicated in the figure. 
As $g_t$ gets smaller, $|p_0^2|$ gets larger and we look at the second interval 
below.

For \intB, 
$|p_0|$ continues to grow, approaching and exceeding $\Lambda_b$. 
The first root is a progressively shallower virtual state 
with $\kappa_1<0$ and $iR_1<0$,
while the second root stays a redundant pole with $\kappa_2>0$ and $iR_2<0$.
Making $Q\to \Lambda_b$ in the relations of the first interval
leads to $a\sim r\sim1/\aleph$.
Numerically, this works well. 
It can be accomplished with 
$\Delta_1\sim \aleph$, 
$\Delta_2\sim \Lambda_b^2/\aleph\gg \Delta_1$,
and $c_1/(2\mu)\sim 1/\aleph$.  
From Eq.~(\ref{eq:AmplitudeEFT}), one 
sees that the second auxiliary field contribution is suppressed by 
$\aleph^2/\Lambda^2$ at small momenta $p\sim\aleph$, and the modified 
ERE increasingly looks similar to the traditional ERE written as a 
Taylor series around $p=0$. 
The situation is depicted in panel (c) in Fig.~\ref{fig:pcotdelta}.
With $c_2/(2\mu)\sim 1/\Lambda_b\ll c_1/(2\mu)$,
the shape-like parameter contribution continues to be suppressed
by $sp^2/r\sim \aleph^3/\Lambda_b^3\ll 1$.

In the third interval, \intC, $|p_0|\sim Q$ becomes very 
large. The fits to \nopi~ scattering phase shift are not sensitive to $p_0$ 
which decouples from the theory. 
The $S$-matrix now has only two poles constrained by $\kappa_1+\kappa_2= 2/r$ 
and $\kappa_1\kappa_2=2/(ar)$,  
 and two residues
 \begin{align}
 |N_1|^2=iR_1=&-\frac{4}{r}\frac{\kappa_1}{\kappa_1-\kappa_2}\, , \nonumber\\
 |N_2|^2=iR_2=&-\frac{4}{r}\frac{\kappa_2}{\kappa_2-\kappa_1}\, . 
 \end{align}
The first root  
continues to get smaller, and at around $g_t\simeq 0.3$ it vanishes. 
Then it moves to the first Riemann energy sheet with a positive normalization 
$iR_1>0$,
signaling the emergence of a bound $nd$ state. 
The second root 
remains a redundant pole, and eventually moves 
slightly beyond the breakdown scale $\Lambda_b$. 
Near the emergence of the shallow bound state, 
the phase shift is characterized by a large scattering length and 
a small effective range, as seen in panel (d) of 
Fig.~\ref{fig:pcotdelta}. 
Qualitatively, the phase shift goes from something similar to the
$^1S_0$ 
2N system with a shallow virtual state to the 
$^3S_1$  
2N channel with a shallow bound state. 
The scattering length scales as $|a|\sim1/\aleph$ whereas the effective range 
$r$ remains fixed at some other small momentum scale set by the 
second root, 
$\kappa_2\sim \aleph' \sim 1/r\gg \aleph$.    
We did not explore how 
$\aleph'$ scales with 
variation of the $nd$ input parameter $a_3$ (through $\Lambda_\star$) in \nopi.   
The halo EFT couplings scale as $\Delta_1\sim\aleph$, 
$c_1/(2\mu)\sim 1/\aleph'$, and $\Delta_2\rightarrow\infty$. 
The second auxiliary field 
is integrated out of the low-momentum theory, and we recover the traditional 
ERE.  
The shape-parameter contribution can be included in the single auxiliary-field 
formulation as a higher-order operator. 
With a scaling $s\sim 1/\Lambda_b^3$, the shape parameter is suppressed by 
$a s p^4\sim \aleph^3/\Lambda_b^3$ compared to the LO scattering-length 
contribution, whereas the effective-range correction is suppressed by 
$a r p^2\sim \aleph/\aleph'$.

\begin{figure}[tb]
\begin{center}
\includegraphics[width=0.49\textwidth,clip=true]{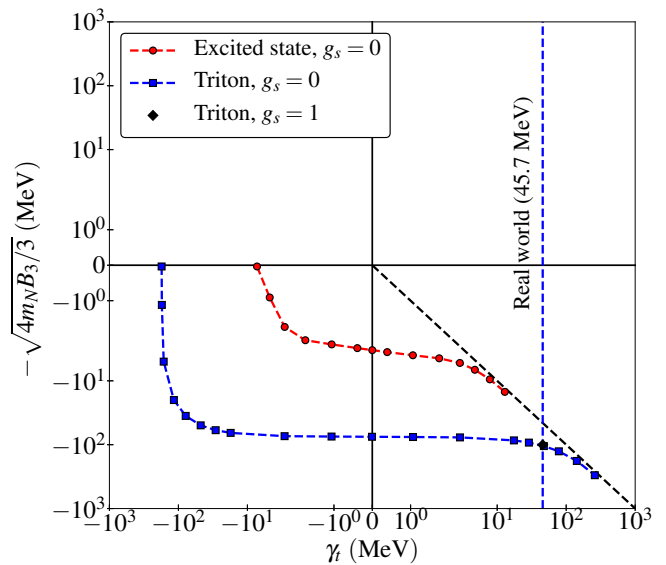} 
\end{center}
\caption{Efimov states: the negative of the three-nucleon binding momentum
($-\sqrt{4m_NB_3/3}$)  
is plotted as a function of the deuteron binding momentum ($\gamma_t$),
both in MeV.
The diagonal dashed line in the fourth quadrant shows the threshold
for breakup into neutron and deuteron.
The physical value of $\gamma_t$ is indicated by the vertical
(blue) dashed line and the triton, by a (black) diamond.
The evolution of the triton at $^1S_0$ unitarity ($g_s=0$) 
is represented by the (blue) squares connected by a (blue) dashed line.
The evolution of the first-excited state is represented by the (red) 
circles connected by a (red) dashed line.
}
\label{fig:Efimov}
\end{figure}

The subsequent evolution of the new bound state is shown in
Fig.~\ref{fig:Efimov}, the ``Efimov plot'' calculated directly
in \nopi.
The physical triton at $g_t=g_s=1$ is seen below the diagonal line on the
fourth quadrant that marks the breakup threshold.
For this bound state,
there is no significant difference between $g_s=0$ and $g_s=1$ 
as the scaling violation due to nonzero $\gamma_s=1/a_s\simeq -8$ MeV 
is a small effect compared to the binding momentum of the triton 
$\sim 100$ MeV \cite{Konig:2015aka}.
As we move towards unitarity with $g_s=0$, $g_t\rightarrow 0$, 
the new, shallow Efimov state appears around $g_t\simeq 0.3$. 
It occurs exactly at the place 
indicated earlier by the halo EFT based on the analytic structure of the 
$S$-matrix. At the unitarity point $g_t=0$, the ratio of binding momenta
between the triton 
and the first-excited state
gives the geometric factor 22.7 predicted by 
Efimov~\cite{Efimov:1970,Efimov:1971,Amado:1971,Amado:1972}. 

Efimov physics displays a limit-cycle 
behavior~\cite{Bedaque:1998kg,Bedaque:1998km}, and shallower bound states
(not shown in Fig.~\ref {fig:Efimov}) also appear.
For example, we have found that, as we evolve towards unitarity, 
around $g_t\simeq 0.05$ a new shallow virtual state 
is present and the phase 
shift goes through the same qualitative behavior as for
the first excited state. Again a modified ERE with 
a new set of scattering parameters 
describes this virtual state, which 
becomes shallower, and finally emerges as the second-excited state
of the triton.
The same process repeats {\it ad infinitum} at progressively smaller
$g_t$ intervals.

\section{Conclusions}
\label{sec:conclusions}

We studied the relation between virtual state and bound Efimov level in a 
three-body system. We chose 
$nd$ scattering in the spin-doublet channel as it 
had been shown to support a virtual state and a bound triton. 
The geometric scaling between bound states had not been observed in this system because 
deeper Efimov levels are  
beyond the range of applicability of any reasonable nuclear theory. 
We have shown that a new shallow Efimov state emerges from the virtual 
state as we drive the system towards unitarity. 
The shallow 
state displays the geometric scaling predicted by Efimov at unitarity,
and we 
find evidence that the accumulation of shallow Efimov levels involve pulling 
in shallow virtual states from the second Riemann energy sheet to the first 
sheet.
Though we consider a specific nuclear system, our results are universal to any three-body
system with resonating zero-ranged two-body interactions, at least one of them supporting a two-body
bound state.
In addition to atomic systems near a Feshbach resonance,
recent lattice QCD calculations, even at unphysical 
quark masses, provide another interesting 
scenario where in the presence 
of a strong magnetic field the  
2N interaction is driven towards unitarity~\cite{Detmold:2015daa}. 

Recently it has been argued \cite{Konig:2016utl} that nuclear ground
states beyond the deuteron are characterized by a momentum scale
intermediate between the pion mass and the inverse 2N scattering lengths.
In this case, nuclei are accessible through \nopi~with
an additional expansion
around the unitarity limit of infinite 2N scattering lengths. 
The existence of a shallow virtual $nd$ state that becomes the triton excited
state supports this picture.

A low-energy halo EFT was formulated for a model-independent description of 
the transition of the shallow virtual to the shallow bound state. We studied 
the analyticity of the $S$-matrix on the complex energy plane 
in order to interpret 
the various poles that correspond to bound, virtual  or redundant states. 
The halo EFT formulated here could be useful in the study of 
low-energy $pd$ scattering,
for example for the model-independent extraction of doublet ERE parameters
from \nopi~\cite{Konig:2014ufa}.
In halo EFT, the Coulomb interaction 
is simpler as $pd$ is effectively a two-body system. 
The halo EFT 
could also be useful in the low-energy 
description of the reactions $d(n,\gamma)^3$H and $d(p,\gamma)^3$He,
which are relevant in big-bang nucleosynthesis. 

\begin{acknowledgments}
The authors acknowledge many helpful conversations with Sebastian K\"onig 
and Jared Vanasse who also provided their triton amplitude 
values for benchmarking our three-body codes, 
and the stimulating environment of the IIP program ``Weakly Bound Exotic Nuclei'' where this project started. 
GR and UvK thank the Institute of Physics at the University of S\~ao Paulo, the 
Institute for Nuclear Theory at the University of Washington, the Kavli 
Institute of Theoretical Physics at the University of California Santa Barbara,
and 
the ExtreMe Matter Institute at the Technical University, Darmstadt, 
for the kind hospitality while this research was carried out.
GR acknowledges partial support from the Joint Institute for Nuclear Physics 
and Applications at Oak Ridge National Laboratory and the Department of 
Physics, University of Tennessee, during his sabbatical where part of this 
research was completed. 
This work was supported in part by FAPESP (RH), project INCT-FNA Proc. No. 464898/2014-5 (RH), 
U.S. NSF grant PHY-1615092 (GR),
the U.S. Department of Energy, Office of Science, Office of Nuclear Physics, 
under award No. DE-FG02-04ER41338 (UvK), 
and the European Union Research and Innovation program 
Horizon 2020 under grant agreement No. 654002 (UvK). 
\end{acknowledgments}

\bibliographystyle{apsrev4-1}
%

\end{document}